\documentstyle[preprint,aps,version2,epsfig]{revtex}
\draft
\def\Journal#1#2#3#4{{#1} {#2} {(#3)} {#4}}

\def\NPA{{ Nucl. Phys. A}}
\def\PLB{{ Phys. Lett.}  B}

\def\PRP{{ Phys. Rep.}}
\def\PRL{ Phys. Rev. Lett.}

\def\PRD{{ Phys. Rev.} D}

\def\EPJ{{Eur. Phys. J.}}

\def\JHEP{J. High Energy Phys.}

\def\ra{\rightarrow}

\def\be{\begin{equation}}
\def\ee{\end{equation}}
\def\bea{\begin{eqnarray}}
\def\eea{\end{eqnarray}}

\def\sbar{{\bar s}}

\def\ANP{{Adv. Nucl. Phys.}}

\begin{document}
\begin{titlepage} 

\bigskip
\begin{center}
{\large\bf 
The quark - antiquark asymmetry of the strange sea 
\\ of the nucleon}

\author{Fu-Guang Cao\thanks{E-mail address: f.g.cao@massey.ac.nz.}
 and A. I. Signal\thanks{E-mail address: a.i.signal@massey.ac.nz.}}
\begin{instit}
Institute of Fundamental Sciences PN461\\ Massey University \\
Private Bag 11 222,  Palmerston North \\
New Zealand
\end{instit}
\end{center}

\begin{abstract}
The strange sea of the proton is generally assumed to have quark - antiquark 
symmetry.
However it has been known for some time that non-perturbative processes involving 
the meson cloud of the proton may break this symmetry.
Recently this has been of interest as it affects the analysis of the so-called 
`NuTeV anomaly', and could explain the large discrepancy between the NuTeV 
measurement of $\sin^{2} \theta_{W}$ and the currently accepted value.
In this paper we re-examine strange - anti-strange asymmetry using the meson 
cloud model.
We calculate contributions to the strange sea arising from fluctuations in the 
proton wavefunction to states containing either Lambda or Sigma hyperons together 
with either Kaons or pseudovector $K^{*}$ mesons. 
We find that we should not ignore fluctuations involving $K^{*}$ mesons in this 
picture. 
The strange sea asymmetry is found to be small, and is unlikely to affect the 
analysis of the Llewellyn-Smith cross section ratios or the Paschos-Wolfenstein 
relationship.

\end{abstract}
\end{titlepage}
\setcounter{footnote}{0}
\section{Introduction}

There has been interest for some time in the question of whether 
non-perturbative processes can lead to a difference between the strange 
and the anti-strange quark distribution functions of the proton. 
This possibility was first pointed out by Signal and Thomas \cite{ST1}, 
and has been subsequently investigated by other authors \cite{BM,CS99,MM}. 
Recently there has been fresh interest in this topic prompted by the 
measurement of $\sin^{2} \theta_{W}$ by the NuTeV collaboration \cite{NuTeV}. 
The large difference between the NuTeV result 
$\sin^{2} \theta_{W}|_{\rm NuTeV} = 0.2277 \pm 0.0013 (stat) \pm 0.0009 (syst)$
and the accepted value $\sin^{2} \theta_{W} = 0.2228 \pm 0.0004$ \cite{SM} 
of around three standard deviations could arise, or be partly explained by, 
a positive value of the second moment of the strange - anti-strange 
distributions $\langle x(s-\bar{s})\rangle=\int_{0}^{1} dx x[s(x) - \bar{s}(x)]$,
as has been pointed out by Davidson and co-workers \cite{n2}. 

As yet there is no direct experimental evidence for any asymmetry in the 
strange sea, however Barone, Pascaud and Zomer \cite{BPZ} found that in 
performing a global fit of unpolarized parton distributions allowing 
$\sbar(x) \neq s(x)$ gave a small improvement to the fit. 
Their best fit result gave the second moment of the asymmetry as 
$\langle x(s - \sbar ) \rangle = 0.002 \pm 0.0028$ at $Q^{2} = 20$ GeV$^{2}$. 
NuTeV have also looked for an asymmetry, and found a small negative value 
for the second moment with a large uncertainty \cite{NuTeV2}. 
However the functional form of the distributions used for this fit was not 
constrained to give the first moment to be zero (i.e. zero net strangeness).

The mechanism for breaking the quark - anti-quark symmetry of the strange 
sea comes from the kaon cloud that accompanies the proton. 
As shown by Sullivan \cite{Sull} in the case of the pion cloud, there is a  
contribution to the parton distributions of the proton from amplitudes where
the virtual photon is scattered from the meson. 
In this case the scaling contribution to the parton distribution of the proton 
can be written as a convolution of the parton distribution of the meson with 
a fluctuation function that describes the momentum probability distribution of 
the meson.
In a similar vein there is a contribution to the parton distribution from 
amplitudes where the virtual photon scatters from the recoil baryon, and the 
meson is a spectator. 
Contributions to the strange sea can come from fluctuations such as 
$p(uud) \rightarrow \Lambda(uds) + K^{+}(u \sbar)$. 
In this case we see that the contribution to the anti-strange distribution, 
which we denote $\delta \sbar$, comes from the anti-strange quark in the kaon, 
whereas the contribution to the strange distribution $\delta s$ comes from the 
strange quark in the Lambda baryon. 
While the valence parton distributions of the kaon and Lambda have not been 
determined by experiment, they can be expected to differ from one another 
considerably, as the $\sbar$ in the kaon carries a larger fraction of the 
4-momentum of its parent hadron than that which is carried by the $s$ quark in 
the $\Lambda$. 
This is certainly the case in comparing the parton distributions of the pion 
with those of the proton, where the pion valence distributions are harder than
the proton valence distributions \cite{Pisf}. 
While the convolution of the meson or baryon valence distribution with the 
appropriate fluctuation function can be expected to decrease the difference 
between them, this difference will lead to a difference between the quark and 
anti-quark distributions in the strange sea of the proton. 

In this letter we re-examine the asymmetry between the strange and anti-strange 
distributions. We work within the context of the meson cloud model (MCM) \cite{MCM}, 
which describes proton $\rightarrow$ meson + baryon fluctuations using an 
effective Lagrangian, with amplitudes calculated using time-ordered 
perturbation theory. 
We consider fluctuations to $\Lambda K^{*}$ and $\Sigma K^{*}$ in addition to the 
$\Lambda K$ and $\Sigma K$ fluctuations of the proton. 
In the non-strange sector, fluctuations involving pseudovector mesons have been 
seen to have significant effects on sea distributions \cite{rho,HHoltmannSS}, 
so we include these fluctuations here to observe whether they make any contribution 
to the asymmetry of the strange sea.

\section{Strangeness in the meson cloud model}

In the meson cloud model (MCM) the nucleon can be viewed as a bare nucleon
plus some meson-baryon Fock states which result from the fluctuation
$N \ra B M$.
The wavefunction of the nucleon can be written as \cite{HHoltmannSS},
\bea
|N\rangle_{\rm physical} =  Z |N\rangle_{\rm bare}
+\sum_{BM} \sum_{\lambda \lambda^\prime} 
\int dy \, d^2 {\bf k}_\perp \, \phi^{\lambda \lambda^\prime}_{BM}(y,k_\perp^2)
\, |B^\lambda(y, {\bf k}_\perp); M^{\lambda^\prime}(1-y,-{\bf k}_\perp)
\rangle 
\label{NMCM}
\eea
where $Z$ is the wave function renormalization constant,
$\phi^{\lambda \lambda^\prime}_{BM}(y,k_\perp^2)$ 
is the wave function of the Fock state containing a baryon ($B$)
with longitudinal momentum fraction $y$, transverse momentum ${\bf k}_\perp$,
and helicity $\lambda$, and a meson ($M$) with momentum fraction $1-y$,
transverse momentum $-{\bf k}_\perp$, and helicity $\lambda^\prime$.
The model assumes that the lifetime of a virtual baryon-meson Fock state is much
longer than the interaction time in the deep inelastic 
process, thus the quark and anti-quark in the virtual meson-baryon Fock states
can contribute to the parton distributions of the nucleon.

For spin independent parton distributions these non-perturbative contributions 
can be expressed as a convolution of fluctuation functions with the 
valence parton distributions in the meson and/or baryon. 
For the strange and anti-strange distributions we have
\bea
x \delta s(x) & = & \int^1_x dy
f_{BM/N} (y) \frac{x}{y} s_{B}(\frac{x}{y}) \, , \nonumber \\
x \delta \sbar(x) & = & \int^1_x dy
f_{MB/N} (y) \frac{x}{y}\sbar_{M}(\frac{x}{y}) 
\label{xqbar}
\eea
where
\bea
f_{BM/N} (y) & = & \sum_{\lambda \lambda^\prime}
\int^\infty_0 d k_\perp^2
\phi^{\lambda \lambda^\prime}_{BM}(y, k_\perp^2)
\phi^{*\,\lambda \lambda^\prime}_{BM}(y, k_\perp^2) \, ,  \\
f_{MB/N}(y) & = & f_{BM/N} (1-y) 
\label{ff}
\eea
are the fluctuation functions. 
These are the probabilities to find the baryon or meson respectively with fraction 
$y$ of the longitudinal momentum.
We note that the relation (\ref{ff}) ensures that while the shapes of 
$\delta s(x)$ and $\delta \sbar(x)$ are different, their integrals are equal, 
ensuring that the strangeness of the dressed nucleon is not changed from that 
of the bare nucleon. 

The fluctuation functions are derived from effective meson-nucleon Lagrangians 
\cite{HHoltmannSS}
\bea
{\cal L}_{NHK}&=&i g_{NHK} {\bar N}\gamma_5\pi H \nonumber \\
{\cal L}_{NHV}&=&g_{NHV} {\bar N} \gamma_\mu\theta^\mu H
			     +f_{NHV} {\bar N} \sigma_{\mu\nu}
		H (\partial^\mu\theta^\nu-\partial^\nu\theta^\mu)
\label{langragians}
\eea
where $N$ and $H$ are spin-1/2 fields, $\pi$ a pseudoscalar field, 
and $\theta$ a vector field.
The fluctuations that we consider are 
$p \rightarrow \Lambda K, \; \Sigma K, \; \Lambda K^{*} \; \mbox{and} \; \Sigma K^{*}$.
The coupling constants that we use are \cite{BHolzenkampHS},
\bea
g_{N\Lambda K}  & = & -13.98,  \nonumber \\
g_{N\Sigma K}  & = & 2.69,  \nonumber \\
g_{N\Lambda K^{*}}  = -5.63, & & 
f_{N\Lambda K^{*}}  = -4.89~{\rm GeV}^{-1}, \nonumber \\
g_{N\Sigma K^{*}}  = -3.25, & & 
f_{N\Sigma K^{*}}  = 2.09~{\rm GeV}^{-1}.
\label{couplingconstant}
\eea
The meson-baryon vertices require form factors, which reflect the fact that the 
hadrons have finite size, and act to suppress large momenta. 
We use exponential form factors 
\bea
G_{MB}(y,k_\perp^2)={\rm exp}\left[\frac{m_N^2-m_{BM}(y,k_\perp^2)}
{2\Lambda_{c}^2}\right],
\eea
though monopole or dipole form factors could also be used with no significant 
difference to our results. 
Here $\Lambda_{c}$ is a cut-off parameter, which appears to be the same for 
fluctuations involving octet baryons and has a value of 1.08 GeV, consistent with 
data on $\Lambda$ production in semi-inclusive $p$-$p$ scattering \cite{HHoltmannSS}. 
Also $m_{BM}^2$ is the invariant mass squared of the $BM$ Fock state,
\bea
m_{BM}^2(y,k_\perp^2)=\frac{m_B^2+k_\perp^2}{y}
+\frac{m_M^2+k_\perp^2}{1-y}.
\eea

In figure 1 we show the four fluctuation functions of interest $f_{\Lambda K /N}(y)$, 
$f_{\Sigma K /N}(y)$, $f_{\Lambda K^{*} /N}(y)$ and $f_{\Sigma K^{*} /N}(y)$ where 
$y$ is the longitudinal momentum fraction of the baryon. 
We note that the kaon fluctuation functions peak around $y = 0.6$, whereas the 
$K^{*}$ fluctuation functions peak around $y = 0.5$ and are fairly symmetrical about 
the peak, indicating that the meson and baryon share the proton momentum equally. 
We also note that the $K^{*}$ fluctuation functions are of similar size to the kaon 
fluctuation functions. 
Thus the higher mass of the $K^{*}$ does not lead to a suppression of these fluctuations, 
as might be expected on kinematic grounds, but to a smaller average meson momentum.
In fact the least probable fluctuation is $p \rightarrow \Sigma K$, which is suppressed 
mainly due to the smaller coupling constant. 
We can conclude that any analysis of strange quark or anti-quarks in the meson cloud 
model needs to include the fluctuations involving the $K^{*}$ meson.

The fluctuation functions depend upon the hardness of the form factor that is used. 
Using a softer form factor (e.g. $\Lambda_{c} = 0.8$ GeV as suggested by reference \cite{MST98}) 
does not change the position of the peaks of the fluctuation functions, but does decreases 
their size. 
The higher mass fluctuations involving $K^{*}$ mesons are more sensitive to the value of 
$\Lambda_{c}$, and decrease in size faster as $\Lambda_{c}$ decreases. 
The first moment of a fluctuation function gives the probability of that particular fluctuation 
occuring. 
For $\Lambda_{c} = 1.08$ GeV we find the total probabilities of finding a kaon or $K^{*}$ to be 
$P(K) = 1.6\% $, $P(K^{*}) = 2.7\%$, whereas for $\Lambda_{c} = 0.80$ GeV these probabilities 
reduce to $P(K) = 0.2\% $, $P(K^{*}) = 0.09\% $.

In order to calculate the MCM contributions $\delta s(x)$ and $\delta \sbar(x)$ 
in eqn.~(\ref{xqbar}) we need to know the parton distributions of the 
$\Lambda$ and $\Sigma$ baryons and the $K$ and $K^{*}$ mesons. 
Previous studies \cite{ST1,CS99,MM} in the MCM have used $SU(3)$ flavour symmetry
to relate these parton distributions to those of the proton and pion. 
However $SU(3)$ flavour symmetry is known to be broken in the sea distributions 
of the proton, so it may not be a good approximation to assume that it holds 
for the valence distributions of octet baryons or the pseudoscalar or pseudovector 
nonet mesons.
A phenomenological parameterization of the kaon distributions, based on those of 
the pion, is available \cite{GRS99}, but there is nothing similar for the baryons. 
An alternative is to use some model of the required distributions. 
One possibility would be to use a gaussian light-cone wavefunction to calculate 
the parton distributions, which is an approach used by Brodsky and Ma \cite{BM}. 
Another approach is to generalise the calculations of the parton distribution 
functions of the nucleon in the MIT bag model by the Adelaide group \cite{ST,SST}. 
This has been done in the case of baryon distributions by Boros and Thomas \cite{BT}, 
and also in the case of the $\rho$ meson by ourselves \cite{CSEPJ}.

Adapting the argument of the Adelaide group, we have the expressions for the 
strange quark distribution of a baryon and the anti-strange quark distribution of 
a meson:
\bea
s_{B}(x) = \frac{m_{B}}{(2\pi)^{3}}  
\int d{\bf p}_{n} \frac{|\phi_{2}({\bf p}_{n})|^{2}}{|\phi_{3}({\bf 0})|^{2}} 
\delta(m_{B}(1-x) - p_{n}^{+}) |\tilde{\Psi}_{+,s}({\bf p}_{n})|^{2}, \\
\sbar_{M}(x) = \frac{m_{M}}{(2\pi)^{3}}  
\int d{\bf p}_{n} \frac{|\phi_{1}({\bf p}_{n})|^{2}}{|\phi_{2}({\bf 0})|^{2}} 
\delta(m_{M}(1-x) - p_{n}^{+}) |\tilde{\bar{\Psi}}_{+,s}({\bf p}_{n})|^{2}.
\label{MITpdf}
\eea
Here we have defined $+$ components of momenta by $p^{+} = p^{0}+p^{3}$, ${\bf p_{n}}$
is the 3-momentum of the 2(1)-quark intermediate state, $\tilde{\Psi}(\tilde{\bar{\Psi}})$ 
is the Fourier transform of the MIT bag wavefunction for the $s$ quark (anti-quark) in 
the ground state, and $\phi_{m}({\bf p})$ is the Fourier transform of the Hill-Wheeler 
overlap function between $m$-quark bag states
\bea
|\phi_{m}({\bf p})|^{2} = \int d{\bf R} e^{-i{\bf p \cdot R}}
\left[ \int d{\bf r} \Psi^{\dagger}({\bf r-R}) \Psi({\bf r}) \right]^{m}.
\eea

The input parameters for the bag model calculations of the parton distributions are 
the bag radius $R$, the mass of the intermediate state $m_{n}$, the mass of the strange 
quark (anti-quark) $m_{s}$ and the bag scale $\mu^{2}$ - at this scale the model is taken 
as a good approximation to ther valence structure of the hadron. 
For the baryons ($\Lambda$, $\Sigma$) we use the same parameter set as Boros and 
Thomas \cite{BT} ($R = 0.8$ fm, $m_{n} = 800$ MeV before hyperfine splitting of 
scalar (for $\Lambda$) and vector (for $\Sigma$) states, $m_{s} = 150$ MeV, 
$\mu^{2} = 0.23$ GeV$^{2}$). 
For the mesons ($K$, $K^{*}$) we use a parameter set based on the baryon set and our 
earlier $\rho$ meson calculation \cite{CSEPJ} ($R =0.7 $ fm, $m_{n} = 425$ MeV, 
$m_{s} = 150$ MeV, $\mu^{2} = 0.23$ GeV$^{2}$).

The valence distributions calculated using eqn. (\ref{MITpdf}) do not satisfy the 
straightforward normalisation condition as they ignore intermediate states with more 
quarks and anti-quarks than the initial state, ie. 3 quarks plus one anti-quark for 
the baryon distribution, or 2 quarks plus one anti-quark for the meson distribution. 
We can parameterise the effects of such states by adding to the distributions a piece 
proportional to $(1-x)^{7}$ for the baryon distributions or $(1-x)^{5}$ for the meson 
distributions, consistent with the Drell-Yan-West relation, such that the normalisation 
condition is satisfied. 
While this ansatz for the shape is somewhat arbitrary, it has little effect at medium 
and large $x$, especially after the distributions are evolved up to experimental scales. 
From the calculated fluctuation functions, we can see that the convolution (equation 
(\ref{xqbar})) is most sensitive to parton distributions in the medium and large $x$ 
regions, which are little affected by our ansatz for the contributions from intermediate 
states with larger mass than the parent hadron.
In figure 2 we show the calculated parton distributions after NLO evolution to 
$Q^2=16$~GeV$^2$, which is the region of the NuTeV data.
We can see that the valance distributions of the $K$ and $K^*$ mesons are harder than
that of the $\Lambda$ and $\Sigma$ baryons, as expected.

Having calculated both the MCM fluctuation functions and the strange valence parton 
distributions of the constituents of the cloud we can now calculate the MCM 
contribution to the proton $s$ and $\sbar$ distributions using the convolution in 
equation (\ref{xqbar}). 
In figure 3 we show our calculated difference between strange and anti-strange parton 
distributions using $\Lambda_c=1.08$ GeV. 
We show the difference calculated with and without including the contributions from 
Fock states involving $K^{*}$ mesons.
We can see that the contributions from $\Lambda K^{*}$ and $\Sigma K^{*}$ are of 
similar magnitude to those from the lower mass Fock states.
As has been noted before \cite{CS02}, the MCM gives a small ($s - \sbar$) difference 
because the harder of the parton distributions ($\sbar^{K}(x)$) is convoluted with the 
softer of the fluctuation functions ($f_{KB/N}$), whereas the the softer parton distribution 
($s^{B}(x)$) is convoluted with the harder fluctuation function ($f_{BK/N}$). 
This leads to similar results for $\delta \sbar$ and $\delta s$, the non-perturbative 
contributions to the parton distributions. 
When $K^{*}$ states are considered, we see from figure 1 that 
$f_{K^{*}B/N}(y) \approx f_{BK^{*}/N}(y)$, so each fluctuation function gives equal 
emphasis to the medium - large $x$ region of the hadron parton distributions when they 
are convoluted together. 
Thus the relative hardness of $\sbar^{K^{*}}(x)$ to $s^{B}(x)$ is manifested in the 
large $x$ region as $\sbar(x) > s(x)$ in the calculation. 
We calculate the second moment of the strange - anti-strange asymmetry both with and 
without $K^{*}$ contributions. 
We find that
\bea
\int_{0}^{1} dx \, x(s(x) - \sbar(x))  =
\left\{
\begin{array}{cc}
1.43 \times 10^{-4} & ~\mbox{without $K^{*}$ states} \\
-1.35 \times 10^{-4} & ~\mbox{including $K^{*}$ states}.
\end{array}
\right.
\eea
We observe that omitting fluctuations involving $K^{*}$ mesons changes the sign of 
the second moment of the asymmetry, so it is important that these fluctuations are 
included in any discussion of the asymmetry of the strange sea. 

Because the calculated asymmetry depends on the difference between MCM contrributions 
to parton distributions, it is relatively insensitive to changes in the form factor 
used to calculate the MCM fluctuation functions. 
If we had used the form factor parameter $\Lambda_{c} = 0.80$ GeV instead of 
$\Lambda_{c} = 1.08$ GeV, we would have obtained values of $4.1 \times 10^{-5}$ 
($3.4 \times 10^{-5}$) for the second moments of the asymmetry without (with) $K^{*}$ 
fluctuations.

The non-zero value of the strange sea asymmetry affects the experimentally determined 
value of the Paschos - Wolfenstein ratio 
\bea
R_{PW} = \frac{\sigma^{\nu}_{NC} - \sigma^{\bar{\nu}}_{NC}}
{\sigma^{\nu}_{CC} - \sigma^{\bar{\nu}}_{CC}} = g_{L}^{2} - g_{R}^{2} = 
\frac{1}{2} - \sin^{2}\theta_{W}. 
\eea
The effect of the strange sea asymmetry is to shift $R_{PW}$ by an amount \cite{n2,CS02}
\bea
\Delta R_{PW} = -\frac{3b_{1}+b_{2}}{\langle x(u_{V}+d_{V})\rangle/2} \langle x(s-\bar{s})\rangle
\label{PWR}
\eea
where 
\bea
b_{1} = \Delta_{u}^{2} = g_{L_{u}}^{2} - g_{R_{u}}^{2};\;\;
b_{2} = \Delta_{d}^{2} = g_{L_{d}}^{2} - g_{R_{d}}^{2}.
\eea
At the NuTeV scale ($Q^{2} = 16$ GeV$^{2}$) the coefficient in front of the second moment 
of the stange sea asymmetry in equation (\ref{PWR}) is  about 1.3, which means that 
$\Delta R_{PW}$ is of the order $1 ~ 2 \times 10^{-4}$.
This is an  order of magnitude too small to have any significant effect on the NuTeV result 
for the weak mixing angle. 

\section{Summary}

Any asymmetry between strange quarks and anti-quarks in the nucleon sea must arise from 
non-perturbative effects. 
This would make any experimental observation of a strange sea asymmetry a crucial test 
for models of nucleon structure. 
We have re-examined this asymmetry within the context of the meson cloud model, which 
gives an asymmetry from strange hadrons in the meson cloud of the proton. 
A novel aspect of our calculation is that we have included the effects of components 
of the meson cloud involving the $K^{*}$ vector meson, and we have seen that the 
contributions to the strange sea from these components are of similar magnitude to 
those involving the pseudoscalar kaon. 
Hence any qualitative discussion of the strange sea in the MCM requires that both sets 
of contributions are considered.
Overall we have found that the strange sea asymmetry in the MCM is fairly small, and
does not have significant effect on the NuTeV extraction of $\sin^{2} \theta_{W}$.
However we have also seen that the sign of the second moment of the asymmetry depends 
on which contributions are considered. 

\section*{Acknowledgments}

We would like to thank Tony Thomas for a number of useful discussions.
This work was partially supported by the Science and Technology Postdoctoral
Fellowship of the Foundation for Research Science and Technology, and the 
Marsden Fund of the Royal Society of New Zealand.

\newpage
\section*{Figure Captions}
\begin{description}
\item
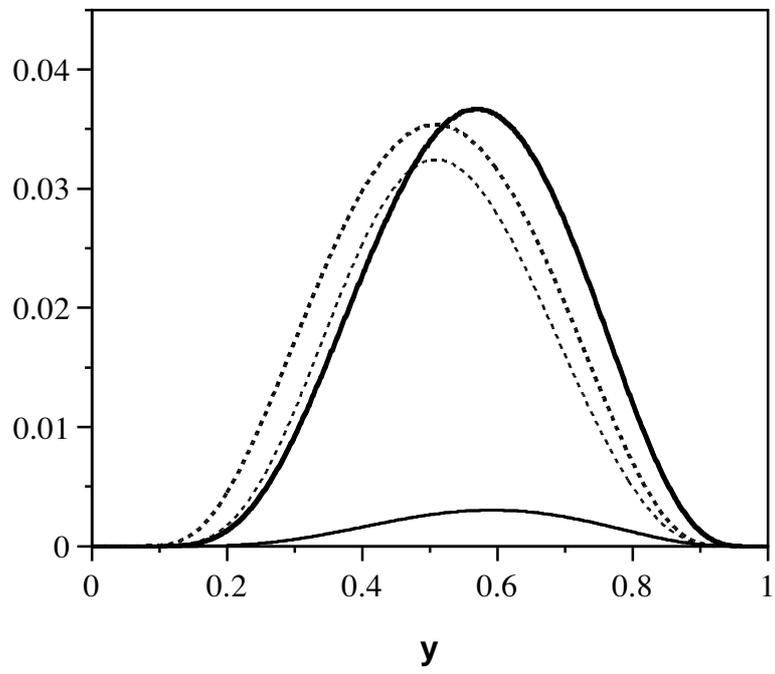
{Fig.~1.} 
The fluctuation functions contributing to the strange sea. 
The thick and thin solid curves are $f_{\Lambda K /N}(y)$ and $f_{\Sigma K /N}(y)$ 
respectively. 
The thick and thin dashed curves are $f_{\Lambda K^{*} /N}(y)$ and 
$f_{\Sigma K^{*} /N}(y)$.
\item
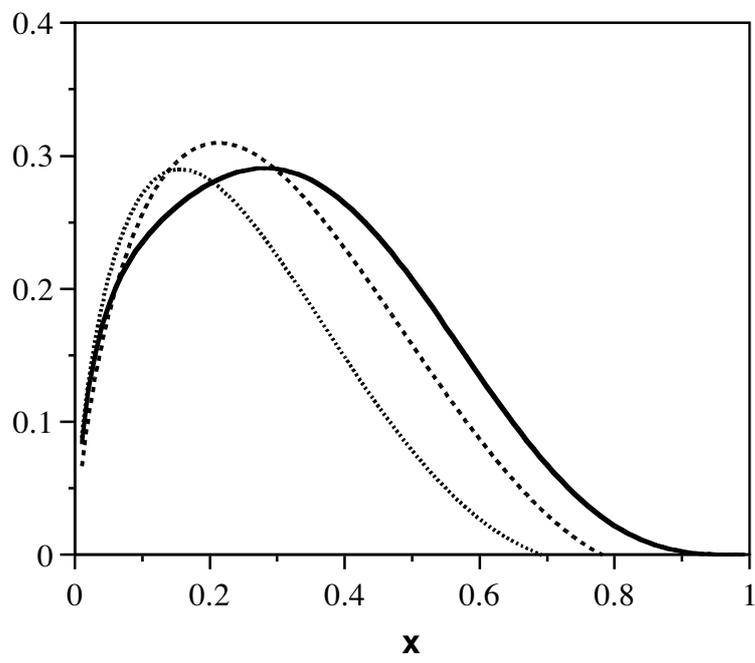
{Fig.~2.}
The valence $\sbar(x)$ distributions of the $K$ and $K^{*}$ mesons (solid curve), and 
the valence $s(x)$ distributions of the $\Lambda$ (dashed curve) and the $\Sigma$ 
(dotted curve) hyperons. 
All distributons are evolved to $Q^2 = 16$~GeV$^2$.
\item
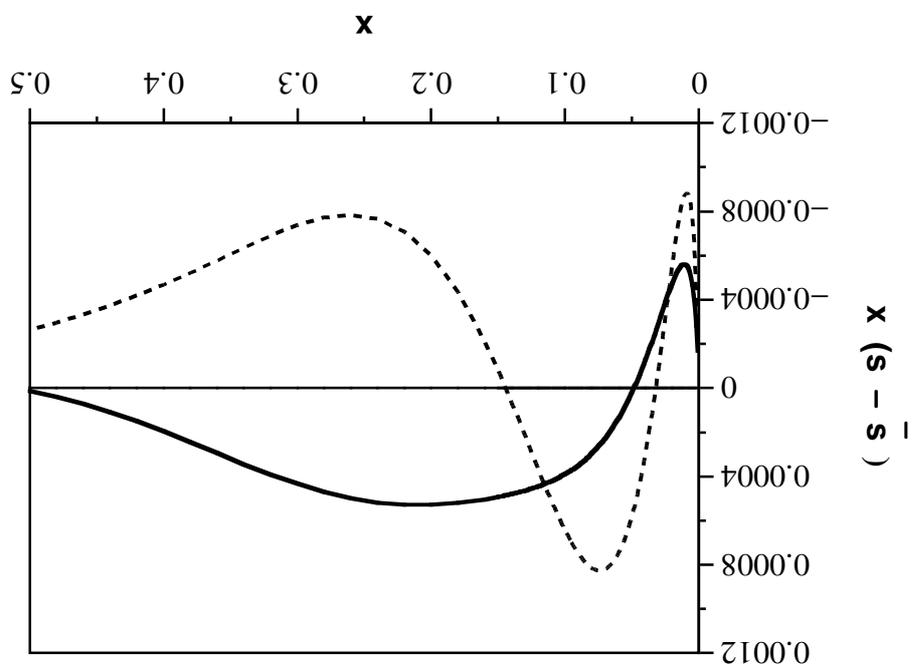
{Fig.~3.}
The strange sea asymmetry $x[s(x) - \sbar(x)]$  calculated in the meson cloud model.
The solid and dashed curves are the results without and with $K^*$ contributions respectively.
\end{description}


\end{document}